\documentclass[conference]{IEEEtran}
 \pdfoutput=1
\usepackage{cite}
\usepackage[pdftex]{graphicx}
 \usepackage{amsmath}
\usepackage{amssymb}
\usepackage{url}

\begin{document}
\title{Measuring electron energy distribution by current fluctuations}
\author{\IEEEauthorblockN{S.U.~Piatrusha\IEEEauthorrefmark{1}\IEEEauthorrefmark{2}, and
		V.S.~Khrapai\IEEEauthorrefmark{1}\IEEEauthorrefmark{2}
}
\IEEEauthorblockA{\IEEEauthorrefmark{1}Institute of Solid State Physics, Russian Academy of Sciences, 142432 Chernogolovka, Russian Federation}
\IEEEauthorblockA{\IEEEauthorrefmark{2}Moscow Institute of Physics and Technology, Dolgoprudny, 141700 Russian Federation}
}
\maketitle
\begin{abstract}
	A recent concept of local noise sensor is extended to measure the energy resolved electronic energy distribution $f(\varepsilon)$ at a given location inside a non-equilibrium normal metal interconnect. A quantitative analysis of $f(\varepsilon)$ is complicated because of a nonlinear
	differential resistance of the noise sensor, represented by a diffusive InAs nanowire. Nevertheless, by comparing the non-equilibrium results with reference equilibrium measurements, we conclude that $f(\varepsilon)$ is indistinguishable from the Fermi distribution.
\end{abstract}

\section{Introduction}

Interest in local measurements of non-equilibrium conductors at the nanoscale motivated various kinds of spatially resolved sensors.
The energy averaged approaches characterize the effective local temperature of the electronic system, via the resistive measurements \cite{Wu2013}, the measurements of tiny heat fluxes \cite{Prokudina2014}, \cite{Menges2016} and the nearfield imaging of a terahertz emission \cite{Weng_Arxiv2016}. Raman thermography permits a local evaluation for the lattice temperature \cite{Doerk2010}. Energy resolved electronic measurements are conventionally based on the energy selective tunneling, which involves a superconducting tunnel probe \cite{Pothier1997} or a quantum dot \cite{Altimiras2010}. In both cases, the energy resolution is naturally limited to the excitations below, respectively, the superconducting gap or dot level spacing.

An alternative approach to gain energy selectivity, without an obviously limited excitation energy, was suggested in 1999 by Gramespacher and B\"{u}ttiker \cite{Gramespacher1999}. They derived a relation between the local electronic energy distribution and the shot noise of a tunneling contact, which served as a bias controlled energy selective probe, see \cite{Meair2011,Kühne2015651} for later developments. Recently, a local noise thermometry was demonstrated by means of diffusive semiconductor nanowires with a resistance much higher than the conductor under test \cite{Tikhonov2016.SciRep}. Here, we extend this approach and perform  energy resolved local noise measurement in a metallic interconnect.

\section{Sensing electron distribution function}

It has been shown recently, that an InAs nanowire (NW) can be used as a miniature noise probe, capable of non-invasive local shot noise measurements in a non-equilibrium conductor \cite{Tikhonov2016.SciRep}. In a system of interest, one end of such nanowire (a test end) contacts a non-equilibrium conductor, while the other (a cold end) is kept at the base temperature $T_0$, thus having an equilibrium electronic energy distribution (EED) at $T_0$. In the case of elastic diffusion, the spatially dependent EED in the NW can be represented as a linear combination of distributions at its ends: $f(x,\varepsilon)= (1-\frac{x}{L})f_{\mathrm{test}}(\varepsilon) + \frac{x}{L}f_{\mathrm{cold}}(\varepsilon)$, where $x$ is the coordinate along the NW \cite{Nagaev1992}. Thus the spectral density of the NW current fluctuations $S_I=\frac{4k_B}{R}T_S$  depends on the EEDs  at both NW ends, since the sensor measures the average noise temperature along the NW: $T_S=\int{T_N(x)\frac{dx}{L}}$, $T_N(x)\equiv\int{f(x,\varepsilon)(1-f(x,\varepsilon))d\varepsilon/k_B}$. For  $T_0 \ll T_N(0)$ this results in $T_S=\alpha T_N(0)$, where $\alpha$ slightly depends on the shape of $f_{\mathrm{test}}(\varepsilon)$ \cite{Sukhorukov1999}.

\begin{figure}[t]
	\begin{center}
		\vspace{0mm}
		\includegraphics[width=0.9\columnwidth]{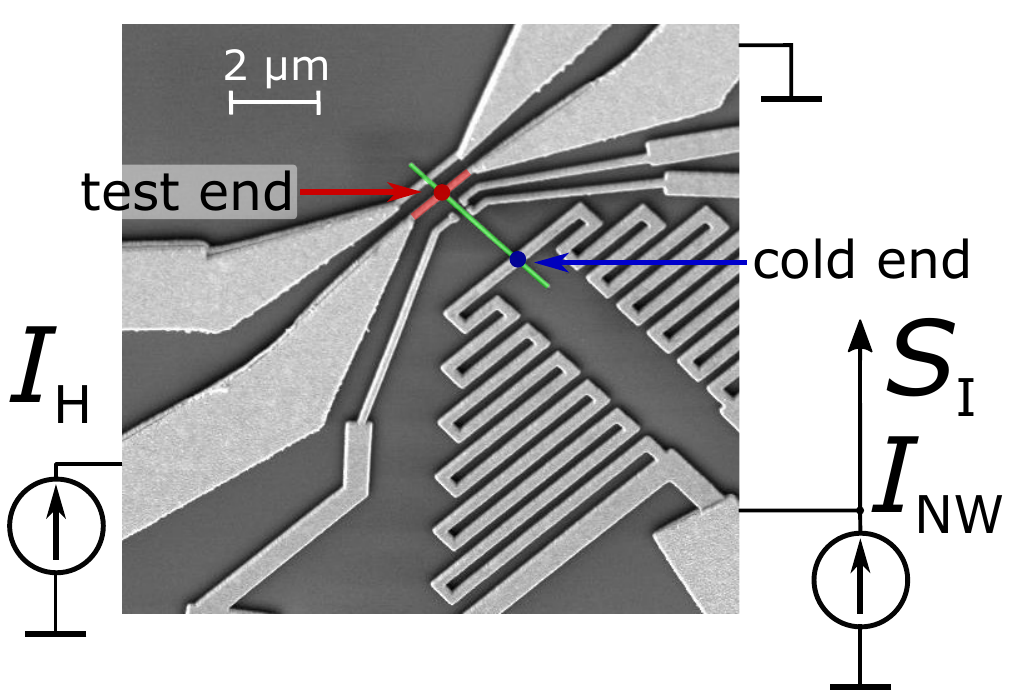}
	\end{center}
	\caption{A sketch of the sample, featuring the golden strip (red) and NW (green). A strip has two current leads, one connected to the current source $I_H$, and other connected to the ground. One end of the NW, denoted as a test end is connected to the center of the golden strip, while another, denoted as a cold end is used to drive current $I_\mathrm{NW}$ through the NW and to measure current fluctuations $S_\mathrm{I}$.
	}\label{fig_sample}
\end{figure}
\par

Here we consider the case, when the electrochemical potential at the cold end can be modified by applying external voltage bias $V_\mathrm{b}$.  In this case, the general equation \cite{Nagaev1992} for current fluctuations $S_I=\frac{4}{RL}\int dx\int d\varepsilon f(x,\varepsilon)(1-f(x,\varepsilon))$ leads to following result:

\begin{equation}
\begin{split}
S_I = \frac{4}{3R} \Big[ &\int d\varepsilon f_{\mathrm{cold}}(\varepsilon)(1-f_{\mathrm{cold}}(\varepsilon)) + \\
&\int d\varepsilon f_{\mathrm{test}}(\varepsilon)(1-f_{\mathrm{test}}(\varepsilon)) + \\
&\frac{1}{2}\int d\varepsilon f_{\mathrm{test}}(\varepsilon)(1-f_{\mathrm{cold}}(\varepsilon)) + \\ &f_{\mathrm{cold}}(\varepsilon)(1-f_{\mathrm{test}}(\varepsilon)) \Big].
\end{split}
\end{equation}

The first two terms represent $T_N(0)$ and $T_N(L)=T_0$ respectively and are independent of  $V_\mathrm{b}$. The other terms, however, contain the product $f_{\mathrm{cold}}(\varepsilon)f_{\mathrm{test}}(\varepsilon)$, which enables the energy selectivity of the noise measurement \cite{Gramespacher1999}. Assuming that $f_{cold}(\varepsilon)=(\exp(\frac{\varepsilon - eV_\mathrm{b}}{k_B T_0})+1)^{-1}$, we obtain the derivative $dS_I/dV_\mathrm{b}$, which simplifies in the limit $T_0 \ll T_N(0)$:

\begin{equation}
\begin{split}
\frac{dS_I}{dV_b} = &\frac{2e}{3R}\int d\varepsilon (1-2 f_{\mathrm{test}}(\varepsilon))\frac{df_{\mathrm{cold}}(\varepsilon)}{d\varepsilon} \\
\approx &\frac{2e}{3R}  (1-2 f_{\mathrm{test}}(\varepsilon=eV_\mathrm{b})).
\end{split}
\end{equation}

Thus, the measured noise and the EED under test are related as:

\begin{equation}
f_{\mathrm{test}}(\varepsilon=eV_\mathrm{b}) = 1/2 - \frac{3R}{4e}\frac{dS_\mathrm{I}}{dV_\mathrm{b}}
\end{equation}

Here, to prove this concept of the EED measurement, we consider a device (figure \ref{fig_sample}) consisting of a short metal strip with the test end of an InAs NW connected to strip's center. The EED $f_{\mathrm{test}}(\varepsilon)$ in the center of the strip is controlled by the external bias current $I_H$. The opposite, cold end of the NW, which is connected to noise measurement circuit is used to apply the bias $V_b$ (hence, the current $I_{\mathrm{NW}}=V_b/R_{NW}$). Other contacts and side gates were not used in the present experiment. This sample has been previously used in \cite{Tikhonov2016.SciRep}, more details on the fabrication can be found in \cite{Tikhonov2016.SST}.

\begin{figure}[t]
	\begin{center}
		\vspace{0mm}
		\includegraphics[width=0.9\columnwidth]{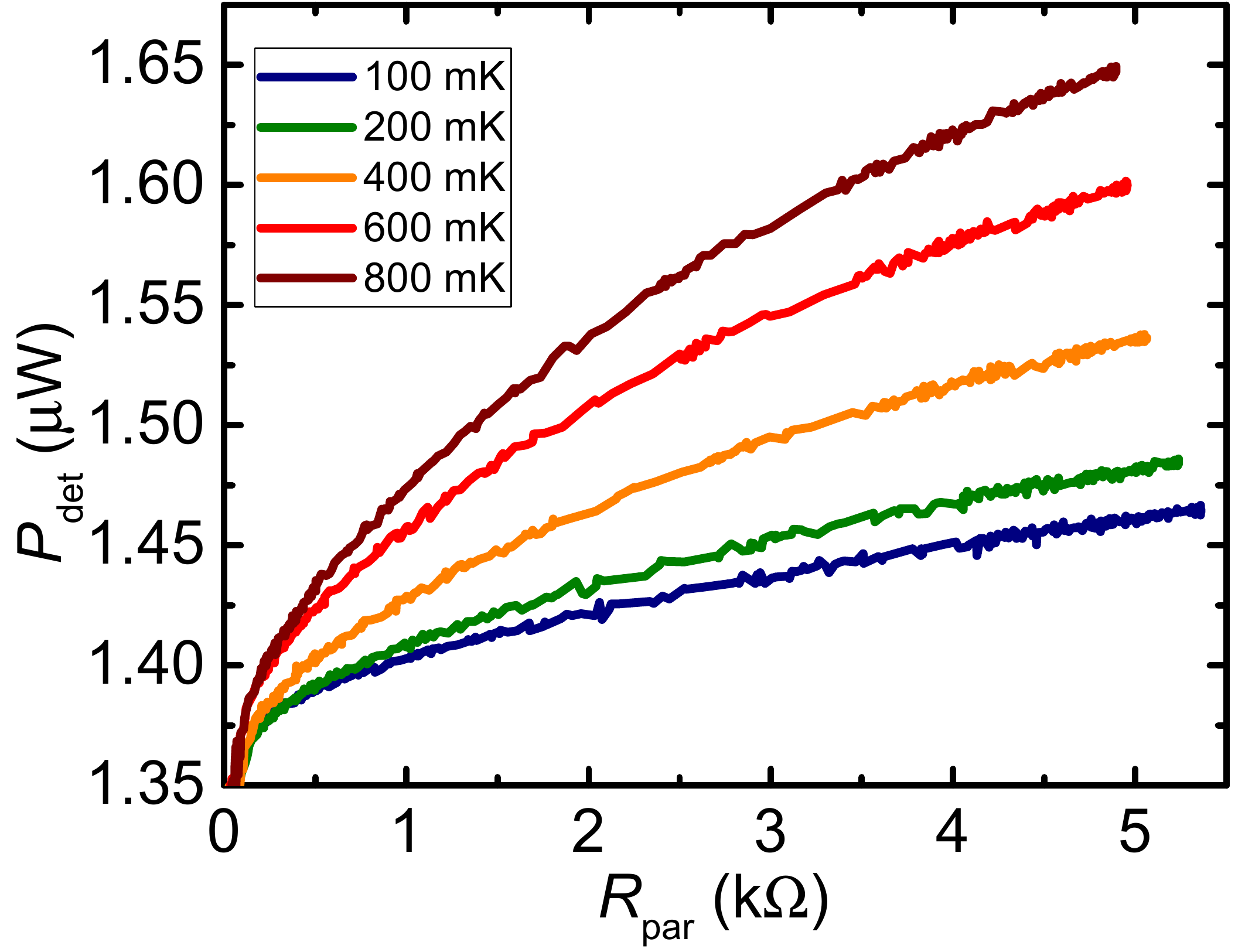}
	\end{center}
	\caption{Power detector response dependence on total noise circuit resistance. Colors correspond to different temperatures. The detector response changes due to change in thermal noise $4k_BT/R$. The resistance-dependent gain coefficient is determined via calibration procedure (see text).
	}\label{fig1}
\end{figure}
\par

\section{Experimental technique}

To measure shot noise, we used resonant amplification of voltage fluctuations from the  cold end, loaded on a $\rm10\,k\Omega$ resistor. Signal is amplified by $\mathrm{\sim75\,dB}$ with an amplifier chain. All measurements were performed in a $\rm ^3He/^4He$ dilution refrigerator with a $\rm30\,mK$ base temperature. A homebuilt low-T amplifier at $\rm\sim800\,mK$ was utilized as a first stage. Noise spectral power density was measured at $\rm\sim8\,MHz$ in a $\rm\sim500\,kHz$ band.

To precisely obtain the unknown full gain coefficient of amplification circuit the Johnson-Nyquist noise calibration procedure was performed. At a given equilibrium $T$, we measured the thermal noise $S_I = 4 k_B T /R_{\mathrm{par}}$ of the sample, the load resistor and the RF transistor, all connected in parallel. The total load resistance was varied between $R_{\mathrm{par}}\approx40\,\rm\Omega$ and $R_{\mathrm{par}}\approx5\,\rm k\Omega$ with the help of transistor gate voltage. 

In Fig \ref{fig1}, we plot the output signal of noise amplification circuit $P_{\mathrm{det}}$ as a function of $R_{par}$ at different $T$. The shape of resulting curves is determined by resistance-dependent full conversion coefficient $G(R_{\mathrm{par}})$, $P_{\mathrm{det}} = G(R_{\mathrm{par}}) S_\mathrm{I}$. This calibration allows to determine both $G(R_{\mathrm{par}})$ and the input current noise of the first stage ($2.5\times10^{-27}\,\rm A^2/Hz$). In addition, we verified that the lowest achievable electronic temperature in our setup is $\rm\approx100\pm20\,$mK.
Throughout the shot-noise measurements the transistor was pinched off.

\section{Results and discussion}

\subsection{Local noise thermometry}

\begin{figure}[t]
	\begin{center}
		\vspace{0mm}
		\includegraphics[width=0.9\columnwidth]{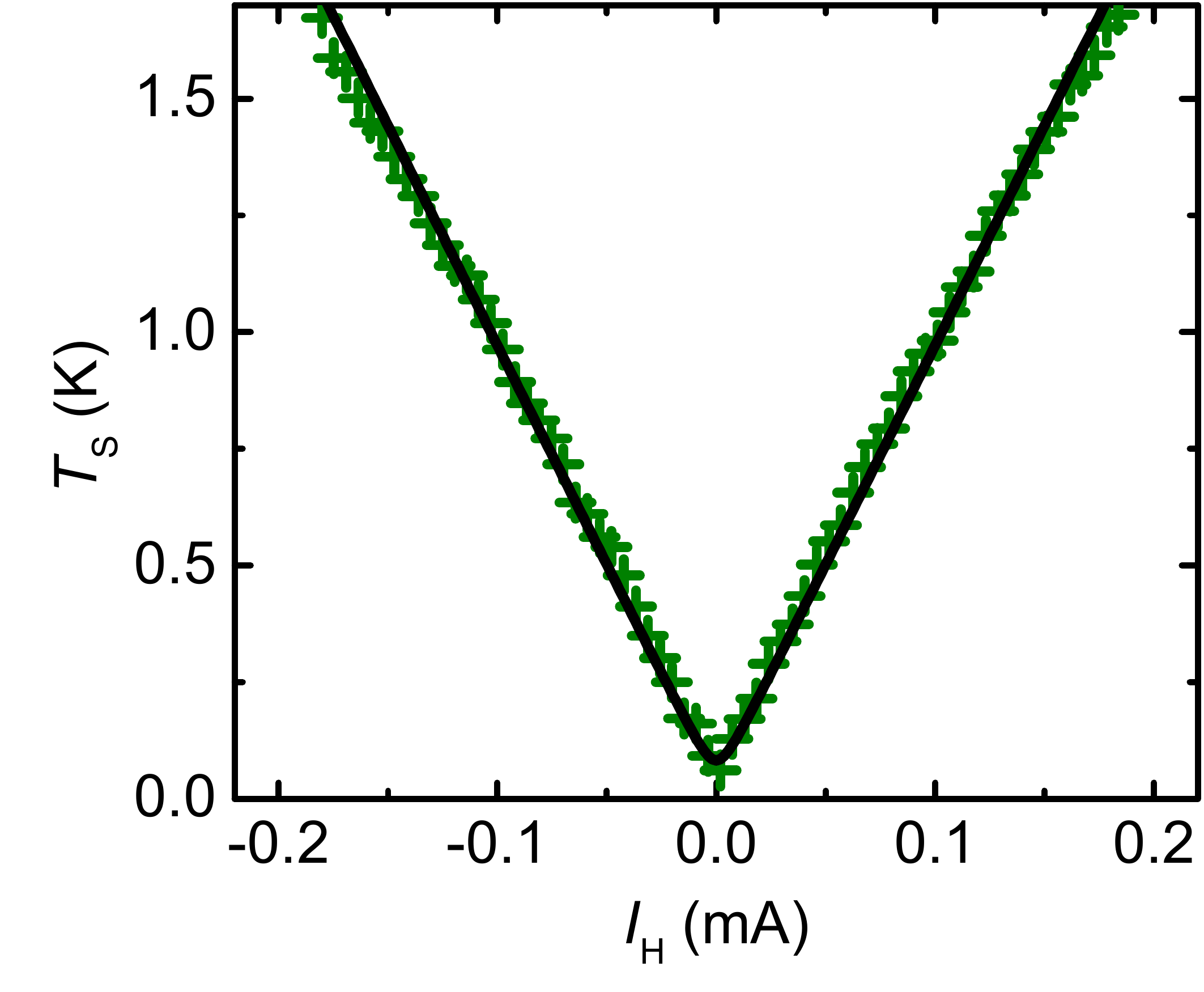}
	\end{center}
	\caption{Three-terminal local noise measurements in the center of a metal strip as a function of the strip current $I_H$. Black line is a shot-noise fit, see text.
	}\label{fig2}
\end{figure}
\par

To test the device operation a local noise measurement with an unbiased NW ($I_{\mathrm{NW}}=0$), similar to \cite{Tikhonov2016.SciRep} was performed at the base temperature. The resulting $T_S$ dependence is shown in figure \ref{fig2} together with a shot noise fit for $T_S$, assuming local equilibrium with temperature:
\begin{equation}
T_N(0)=\sqrt{T_0^2 + \frac{3}{4\pi^2}(e r I_H/k_B)^2},
\label{eq_balance}
\end{equation}
determined by balance between Joule heating and Wiedemann-Franz heat conductance \cite{Nagaev1995}. The strip resistance fit parameter equals $r = \rm5.2\,\Omega$, substantially higher that the value $r \sim \rm3\,\Omega$ previously obtained in \cite{Tikhonov2016.SciRep}. Most probably, this is explained by the fact, that in previous experiments the sample was immersed into liquid helium, which resulted in better thermalization of the current leads. 

\begin{figure}[t]
	\begin{center}
		\vspace{0mm}
		\includegraphics[width=0.9\columnwidth]{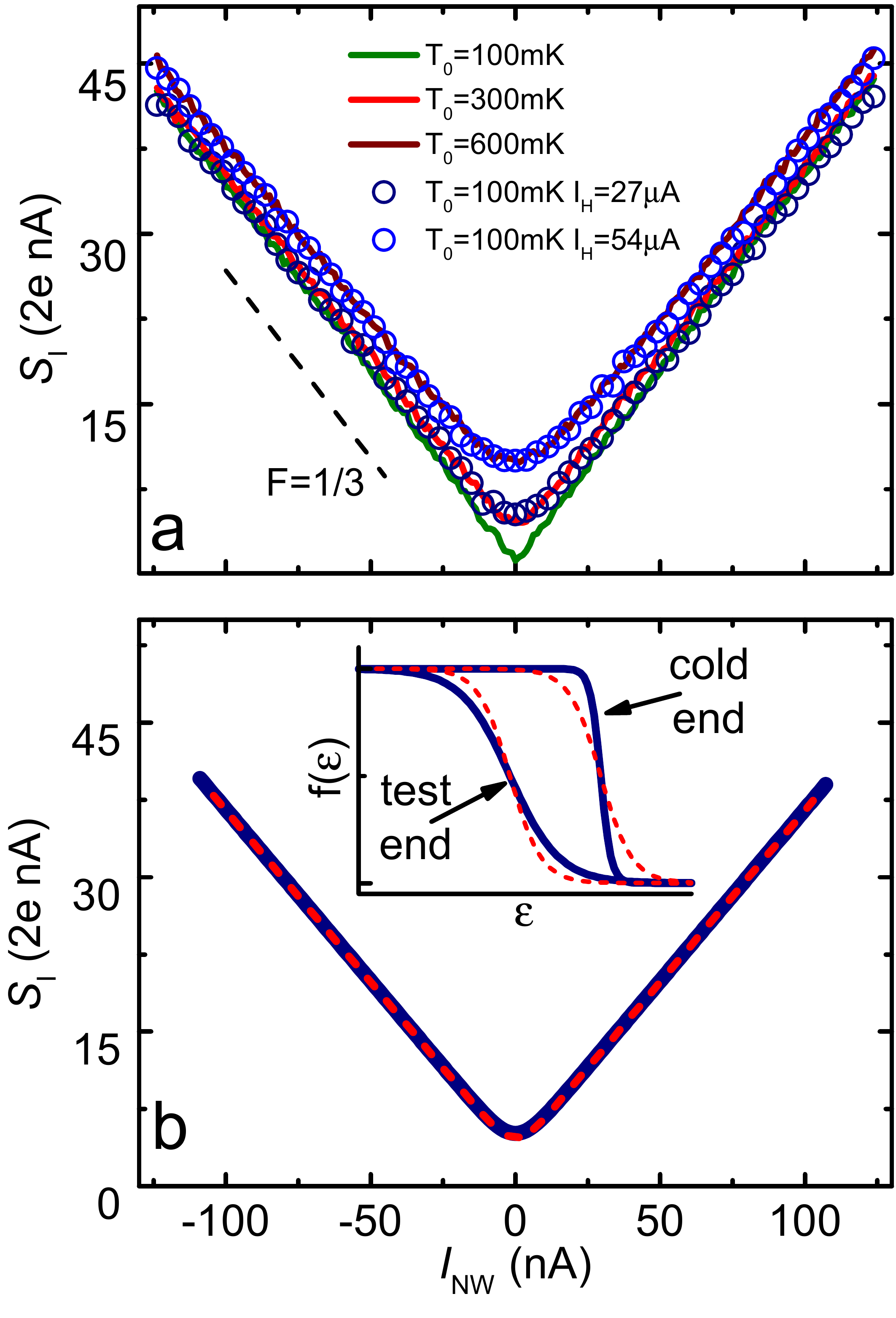}
	\end{center}
	\caption{(a) The experimentally measured current noise as a function of NW current. Solid lines represent equilibrium configuration, when no current flows through metal the strip at different temperatures. Symbols represent non-equilibrium configurations with different strip currents. Dashed line shows Fano factor $\rm1/3$ slope.
		(b) Theoretical prediction of the current noise for equilibrium $\rm300\,mK$ case (red dashed line) and non-equilibrium $\rm100\,mK/27\,\mu A$ case (dark blue solid line). Inset: a sketch of EEDs at test and cold end for equilibrium and non-equilibrium cases at an equal $I_{\mathrm{NW}}$. Solid blue lines for non-equilibrium case are Fermi-Dirac distributions with different temperatures and chemical potential offset, while red dashed lines for equilibrium case are Fermi-Dirac distributions with the same temperature differing only by offset.
	}\label{fig3}
\end{figure}
\par

\subsection{Electronic energy distribution}

To verify the concept of the EED measurement the shot noise dependence on $I_{NW}$  was measured in two configurations: equilibrium strip ($I_H=0$ and $T_0$ above the base) and non-equilibrium strip ($I_H\neq0$ and base $T_0$). The $T_0$ in the first case and the $I_H$ in the second case were adjusted such that in the absence of the NW bias ($I_{\mathrm{NW}}=0$) the $T_S$ is the same in two experiments. 

The green curve in figure \ref{fig3}a corresponds to the standard shot noise measurement at the base temperature and $I_H=0$. The Fano factor $F\approx1/3$ proves the elastic diffusive transport in the NW \cite{Beenakker_Buettiker_1992}\cite{Nagaev1992}, which is crucial for operating the noise sensor.

\begin{figure}[t]
	\begin{center}
		\vspace{0mm}
		\includegraphics[width=0.9\columnwidth]{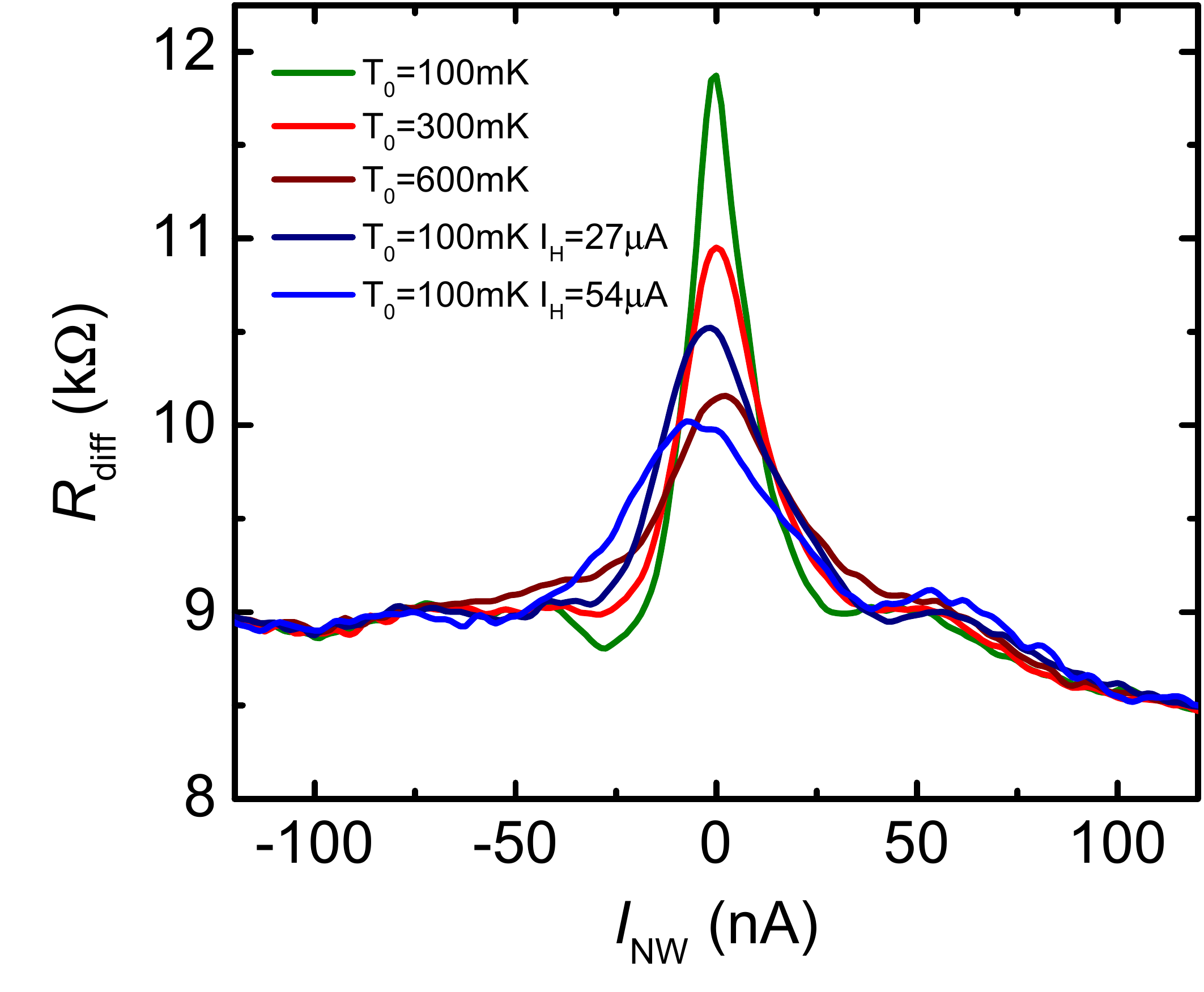}
	\end{center}
	\caption{NW differential resistance $R_{\mathrm{diff}}=dV/dI$ as a function of NW current I at different overheat regimes.
	}\label{fig_Rdiff}
\end{figure}
\par

At increasing $I_H$ (symbols) or $T_0$ (lines) the measured noise increases, as expected. However, the noise in the non-equilibrium case is indistinguishable from equilibrium case with higher corresponding $T_0$ (red line with dark blue symbols, and dark red line with blue symbols). To verify if this observation is consistent with the EED sensing we plot the theoretical predictions for current noise in corresponding configurations, see figure \ref{fig3}b. For the non-equilibrium case, the electron temperature at the test end was calculated using eq. (\ref{eq_balance}). Similar to the experimental data, the results for the equilibrium and non-equilibrium cases are also almost indistinguishable in figure \ref{fig3}b. We conclude that the EED in the middle of the current biased strip is very well captured by the Fermi-Dirac distribution with the local temperature given by eq. (\ref{eq_balance}). A direct comparison to the experimental data is complicated because of a slightly nonlinear current-voltage response of the NW, which gives rise to $I_{\mathrm{NW}}$ dependent differential resistance (see figure \ref{fig_Rdiff}). Note that in spite of this similarity, the case of non-equilibrium strip is characterized by a strong temperature gradient along the NW, which manifests itself in thermoelectric measurements \cite{Tikhonov2016.SST}. The EEDs on the two ends of the NW are sketched in the inset of figure \ref{fig3}b.

\section{Conclusion}
In summary, we experimentally realized the concept of the energy selective local noise measurement. The nonlinear current-voltage response of the NW complicates accurate extraction of the local EED under test. Yet, comparison with the theoretical calculations is consistent with the Fermi-Dirac shaped EED. 

\section*{Acknowledgment}
We thank D.V.\,Shovkun  and E.S\, Tikhonov for fruitful discussions. We gratefully acknowledge S. Roddaro and L. Sorba for providing us with the NW devices. This work was supported by the Russian Science Foundation under the grant RSF-DFG: 16-42-01050 and the bilateral CNR-RFBR project 15-52-78023.

\bibliographystyle{IEEEtran}
\bibliography{nwdistrbbl}

\end{document}